\documentstyle[12pt,aasms4,flushrt]{article}

\begin{document}
\title{An Excess of C {\sc iv} Absorbers in Luminous QSOs:\\
 Evidence for Gravitational Lensing?}
\author{Daniel E. Vanden Berk, Jean M. Quashnock, Donald G. York}
\affil{University of Chicago\\ Dept. of Astronomy and Astrophysics\\
 5640 S. Ellis, Chicago, IL 60637}
\author{Brian Yanny}
\affil{Fermi National Accelerator Laboratory\\ 500 Wilson Road, MS 127\\
 Batavia, IL 60510}
\authoremail{devb@oddjob.uchicago.edu}
\authoremail{jmq@oddjob.uchicago.edu}
\authoremail{don@oddjob.uchicago.edu}
\authoremail{yanny@sdss.fnal.gov}

\slugcomment{To appear in {\it The Astrophysical Journal}, 1996 September 20.}

\begin{abstract}
We have compiled a new and extensive catalog of heavy-element QSO absorption
line systems and analyzed the distribution of absorbers in bright and faint
QSOs, to search for gravitational lensing of background QSOs by the matter
associated with the absorbers. There is a highly significant excess of C {\sc
iv}  absorbers in bright QSOs in the redshift range $z=1.2-3.2$, and this
excess increases strongly as a function of QSO absolute magnitude. No
significant excess is found for Mg {\sc ii} absorbers in the redshift range
$z=0.30-1.55$.  We rule out several possible reasons for this effect and
argue that the C {\sc iv} excess could be due to  gravitational lensing.  If
so, then the lensing masses must be at $z \gtrsim 1.5$ and within several
hundred comoving Mpc of the QSOs, where the C {\sc iv} absorbers are mainly
found.  The absence of an excess in the available Mg {\sc ii} sample would
then arise because the Mg {\sc ii} data does not sample this region of space. 
\end{abstract}


\section{Introduction}
One possibly important bias in the study of QSO absorption line systems is
that the observed number density of absorbers may be higher than the true
number density because of gravitational lensing of the backlighting QSOs.
Since there is an observational bias at all redshifts towards QSOs with 
brighter apparent magnitudes, the current sample of QSOs searched for 
absorption line systems may contain a large fraction of QSOs observed only 
because they have been gravitationally brightened beyond some apparent 
magnitude limit.

Significant lensing mass may be associated with a higher than average
density of galaxies and thus a higher than average number of QSO absorbers.
This possible association could either be one-to-one, with
the lensing mass identified with the absorber (Bartelmann \& Loeb 1996), or
simply statistical, where the number of absorbers is correlated with the
lensing mass.  In either case, the geodesics to the QSOs no longer represent
random lines-of-sight through the universe with respect to the counts of
absorbers.
 
If this bias were sufficiently strong, the current interpretation of absorption
line number statistics would have to be reconsidered.  However, it might assist 
in the search for gravitationally lensed QSOs.
For example, lensed QSO candidates could be identified by searching for those 
QSOs that are both apparently very luminous and have a significant excess of 
absorption line systems.  This bias would also directly affect studies of the 
QSO luminosity function, since it would be difficult to disentangle the 
effects of QSO evolution from gravitational lensing without some observational
signature of lensing other than luminosity (cf.\ Pei, 1995).

Several attempts have been made to determine if the lensing bias exists and
to assess its importance in studies of absorption line statistics. Thomas
and Webster (1990) examined the possibility of lensing by compact objects
associated with QSO absorbers for several homogeneous absorber samples. 
After attempting to fit several lensing models to the absorber redshift
distribution, they concluded that contamination of the samples by absorbers
associated with compact lenses is likely to be small for both C {\sc iv}
and Mg {\sc ii} systems; however, lensing associated with some absorbers,
especially ones with high equivalent widths, could not be ruled out. 
Steidel and Sargent (1992) applied the same test to their large homogeneous
sample of Mg {\sc ii} absorbers, and concluded that at most a few percent
of the QSOs in their sample were observed only because they were brightened
by absorbers associated with lensing systems.

Our aim here is not to determine a precise lensing model that may be
consistent with the data, but to see if there is any evidence that
lens-associated absorption is common. This approach was taken by York et
al.\ (1991, hereafter Y91), who looked  at the redshift number density of
absorbers in bright versus faint QSO samples in a large heterogeneous
collection of heavy-element absorption line systems.  They found a factor
of two increase in the number of absorbers per unit redshift in the bright
sample over the faint sample at high ($z \sim 3$) redshifts, and no excess of
absorbers between $0.4<z<2$.\footnote{We note that there was a
computational error in the calculations of the redshift number density of
Y91; however, the corrected result of the lensing examination remains
essentially unchanged.} We re-examine this effect using a new and
more extensive sample of absorbers, and we present several other tests
which show evidence for lens-associated absorption in this sample.

\section{The Absorber Catalog and Selection of the QSO and Absorber Samples}
The catalog of QSO absorption line systems contains data on all
heavy-element absorption lines in the literature, complete up to October
1994, with some additional entries since then.  It is an updated version of
the QSOALS catalog of Y91, but is more than twice the size, with over 2200
absorbers listed in 484 QSOs, and is the largest sample of heavy-element
absorbers compiled to date.  More details on the catalog in general can be
found in Y91. We have eliminated 7 QSOs with multiple images that are known
or strongly suspected to be gravitationally lensed: 0142-100, 0957+561,
1115+080, 1208+101, 1413+117, 1634+267, and 2345+0007.  It is not clear if
the addition of these QSOs would, in principle, bias our sample one way or
the other. The results in the following sections are virtually unaffected
when they are included.

Although the catalog of absorbers is compiled from data taken with a wide
variety of telescopes and spectrographs, and taken under a range of
observing conditions, for various purposes, the information included is
enough, in principle, to construct a homogeneous sample of heavy-element
absorption line systems.  In particular, we need to know the wavelength
limits and equivalent width limits (which take into account signal-to-noise
and resolution) for each QSO spectrum, in addition to the redshifts $z$,
equivalent widths {\sl W}, and line identifications for each absorber. 
The observed spectral equivalent width limit will generally be a function of
wavelength, but the catalog lists only one 1$\sigma$ estimate for the entire
wavelength range of each spectrum.  Partly to overcome this deficit, and
partly because different authors use different equivalent width criteria for
including absorption lines in their lists, a conservative 5$\sigma$
observed equivalent width limit for each spectrum was used to select
absorbers from the catalog.

Selected absorption systems must have been identified with at least
three lines, or two lines if they belonged to a doublet, in which case the
equivalent width ratio of the intrinsically weaker to stronger doublet member
must be less than or equal to 1, within the listed measurement errors.  We
have concentrated on absorbers containing either C {\sc iv} 1548{\AA} or
Mg {\sc ii} 2796{\AA}, since these are the most commonly identified
absorption lines. Since absorbers with these lines are usually seen in
different redshift regimes, each ion has been
treated separately in the following analysis.  Only a small number of
absorbers are observed with both C {\sc iv} and Mg {\sc ii} lines. We have
analyzed several other lines identified in relatively large numbers
including Si {\sc ii} 1527{\AA}, which
will be discussed in the context of the C {\sc iv} and Mg
{\sc ii} results in Section 4.  We note that there are only 42 (out of 655)
reliably identified  absorbers in the catalog in which C {\sc iv} or Mg
{\sc ii} was not detected even though their lines would have fallen into the
observed spectral range.

The Ly $\alpha$ forest region of each spectrum has been excluded, since
it is generally very difficult to unambiguously distinguish lines of
heavy-element ions from the more numerous Ly $\alpha$ lines. The entire
spectral range within 5000 km/s of the QSO emission and the absorbers
identified there have been excluded, as is commonly done, since there may be
a direct association between the QSOs and the absorbers in that region
(Aldcroft, Bechtold, \& Elvis, 1994).
QSO spectra with very poor ($>400$ km/s) and very good ($<35$ km/s) velocity
resolution have also been eliminated to avoid blending of distinct
absorbers in the former case, and to avoid splitting single absorbers into
multiple components in the latter case, either of which could distort the
counts of absorbers.  After these selection criteria were applied to the
catalog, 398 C {\sc iv} and 257 Mg {\sc ii} absorbers remained.  In the
following analysis, we have not combined absorbers that may originate in the
same galaxy ($\Delta v \lesssim 500$ km/s), but doing so does not
significantly affect the results.

\section{Analysis of the QSO and Absorber Samples}

We now examine correlations of absorber properties with estimated QSO 
luminosity.  We find no significant correlations, in either the C {\sc iv} 
or Mg {\sc ii} absorber samples, between QSO absolute magnitude, 
M$_{\rm V}$, and either
absorber redshift, absorber equivalent width, observed spectral equivalent
width limit, or spectral resolution, except possibly at the extreme ends of
the QSO absolute magnitude distribution which have been excluded as 
described below.

Our first test for lensing associated with absorbers is to compare the 
redshift number density of absorbers, $d{\cal N}/dz$, in ``bright'' and
``faint'' samples of QSOs, similar to the lensing test described in Y91.
Although the sample of QSOs, spectra, and absorbers has been selected
with a set of uniformly applied criteria, an estimate of $d{\cal N}/dz$
must take into account the remaining variations in wavelength ranges, observed
equivalent width limits, and absorber rest equivalent widths.  To do this,
we have used the following  weighting scheme for the absorbers.

Each detected absorption line occupies a point in the $z$-{\sl W} plane,
and each QSO spectrum samples some $z$ and {\sl W} region in
that plane according to the selection criteria referred to above.  
The density of absorbers per QSO,
$\partial^{2} {\cal N}/{\partial z \partial {\sl W}}$,
is a series of delta functions centered at each $(z,{\sl W})$ absorber point,
weighted by the reciprocal of $S_a$, 
the number of QSO spectra in which an absorber
{\it could} have been detected in both $z$ and {\sl W},
\begin{equation}
\partial^{2} {\cal N}/{\partial z \partial {\sl W}}=\sum_{a}S_{a}^{-1}\delta
({\sl W}-{\sl W}_{a})\delta(z-z_{a}).
\end{equation}
The average number of absorbers per QSO, ${\cal N}$, in some redshift and
equivalent width region is the integral of 
$\partial^{2} {\cal N}/{\partial z \partial {\sl W}}$,
which is just the sum of the weights in that region. 
Then in the ranges $\Delta z$ and $\Delta {\sl W}$, the average redshift
number density will be $d{\cal N}/dz={\cal N} / \Delta z$, and the average
equivalent width number density will be $d{\cal N}/d{\sl W}={\cal N} / 
\Delta {\sl W}$.

Cuts in equivalent width and redshift are only necessary in regions that are
poorly sampled, i.e., where $S_a$ is small.  The well-sampled regions of
the $z$-{\sl W} plane for C {\sc iv} and Mg {\sc ii} are bounded by the limits
shown in Figures 1 and 2. Mg {\sc ii} absorption has often been observed at
higher redshifts, but the high$-z$ cutoff has been set to $z=1.55$ because
telluric absorption and night sky emission lines can reduce the reliability
of Mg {\sc ii} detections beyond that point.

The QSOs have been divided into bright and faint samples in two ways. First,
the median value of the QSO absolute visual magnitude distribution (assuming
$q_{o}=1/2$, $H_{o}=100$h km s$^{-1}$ Mpc$^{-1}$, and $\alpha=0.7$
(Veron-C\'{e}tty \& Veron 1991)) was used as the dividing point for the
bright and faint QSOs, after cuts in M$_{\rm V}$ (seen as the limits in
Figure 4) were made to ensure adequate sampling of $z$ and {\sl W}.  The
sample division is nearly independent of $q_o$.  Second, we fit a line
through a plot of QSO redshift versus apparent visual magnitude, so that at
each redshift there are roughly equal numbers of QSOs above and below the
line.  Here, both QSO luminosity evolution, and any selection effects where
more luminous QSOs are selected at higher redshifts, are taken into account by
defining a bright and faint sample at each redshift.  We have found that
the two methods divide the QSO samples roughly the same way, except at the
lower redshifts where the slope in the M$_{\rm V}$ distribution versus
redshift is fairly steep, resulting in more QSOs in the faint half of the
M$_{\rm V}$ division at lower redshifts. The results for both methods are
presented here.

Figure 1 shows Mg {\sc ii} and C {\sc iv} $d{\cal N}/dz$ versus redshift for
the bright and faint QSO samples. The solid histograms show the average
$d{\cal N}/dz$ for 100 randomly selected QSO samples of the same sizes as
the real samples, and the 1$\sigma$ deviations in those samples are shown
by the dashed histograms.  Figure 1 clearly shows that bright QSOs in the
C {\sc iv} sample have more absorbers per unit redshift than the faint
sample, and that this seems to be true at all the observed redshifts. The
probability that the observed C {\sc iv} $d{\cal N}/dz$ difference in
the bright and faint samples is the result of chance is $7.4 \times
10^{-6}$ for the division of QSOs by M$_{\rm V}$, and $6.0 \times 10^{-11}$
for the line fit division. There is no significant difference, however,
between the bright and faint halves of the Mg {\sc ii} sample (probabilities
0.16 and $7.9 \times 10^{-2}$ for the respective divisions). 

The difference seen in the C {\sc iv} sample is {\it not}  due to a higher
density of small equivalent width systems in the bright QSO sample. Fig.\
2 shows $d{\cal N}/d{\sl W}$ versus absorber equivalent width.  While the
{\it amplitude} of $d{\cal N}/d{\sl W}$ is higher for the bright C {\sc iv} 
sample than for the faint (as expected from the $d{\cal N}/dz$ result), the 
{\it shapes} of the $d{\cal N}/d{\sl W}$ distributions are virtually the same.
The excess occurs at {\it all} values of ${\sl W}$. Thus the absorbers in both
the bright and faint QSO samples seem to have identical equivalent width 
properties, except that there are more absorbers in the bright sample.  
Figure 2 shows that the Mg {\sc ii} absorbers also have a similar equivalent
width distribution in the bright and faint halves, but as expected from the 
Mg {\sc ii} $d{\cal N}/dz$, the amplitudes of both $d{\cal N}/d{\sl W}$ 
distributions are about the same. 

Systematic differences between observers in defining signal-to-noise in 
their spectra, combined with an uneven distribution of bright and faint QSOs
among the observers for the C {\sc iv} sample only, do not explain the noted
effects.  We show in Fig.\ 3 the C {\sc iv} $d{\cal N}/dz$ and
$d{\cal N}/d{\sl W}$ distributions in
bright and faint QSO samples (using the median M$_{\rm V}$ as the dividing
point) for the single largest homogeneous C {\sc iv} sample (Sargent,
Boksenberg, \& Steidel 1988; Steidel 1990) which includes 144 absorbers in
57 QSOs which have available visual magnitude estimates.  These samples include
the ``complete sample'' of systems with ${\sl W}\geq 0.15${\AA} described in
Steidel (1990), excluding those systems within 5000 km/s of the QSO emission.
As in Fig.\ 1, the value of $d{\cal N}/dz$ in the brighter half is
always higher than in the fainter half.  The probability of the observed
difference occuring by chance is $6.6 \times 10^{-4}$.
The slopes of the $d{\cal N}/d{\sl W}$
distributions in Fig.\ 3 are almost identical, while the amplitudes differ 
by a nearly fixed amount, as is the case with the C {\sc iv} distributions in
Fig.\ 2.  Thus the homogeneous sample shows the same behavior as the full
sample, and inhomogeneity is not the cause of the observed C {\sc iv} excess.

A second test for lensing is to compare the number of absorbers seen at each
QSO absolute magnitude with the average number that would be expected at
that magnitude.  If lens-associated absorption occurs, we expect to find
more absorbers on average in the spectra of more luminous QSOs, because the
fraction of lensed QSOs should increase as a function of luminosity. The
average number of absorbers expected in a QSO, ${\cal N}$, can be found by
summing the weights in equation 1 that lie in the region of the $z$-{\sl W}
plane covered by the QSO spectrum. Fig.\ 4 shows the total number of
absorbers observed ${\cal N}_{obs}$, divided by the average expected total
number of absorbers ${\cal N}_{ave}$, for each M$_{\rm V}$ bin.  There is a
dramatically strong correlation ($Q(\chi^{2})=4\times 10^{-4}$)
between the number of absorbers and M$_{\rm
V}$ for the C {\sc iv} sample, while there is no significant trend in the
Mg {\sc ii} data. The absence of a correlation in the Mg {\sc ii} data
supports the argument that the C {\sc iv} results are not due to any
selection effects since both samples come from the same QSO spectra and
were analyzed in precisely the same way. Both of the results in Fig.\ 4 are 
consistent with the $d{\cal N}/dz$ distributions for the bright and faint QSO 
samples.  For the homogeneous C {\sc iv} sample described above, the results 
are consistent with Fig.\ 4 but are less significant owing to the smaller
QSO sample size.

\section{Discussion}

We find clear evidence for an 
excess of C {\sc iv} absorbers in bright QSOs.  We have allowed
for differences among the various data sets in the catalog by carefully
comparing equivalent width detection limits, resolution, and observable
redshift limits.  For reasons discussed above, we find it improbable that
selection effects are causing the observed results.  First, the
effect occurs at all values of ${\sl W}$, and is not due to higher spectral
sensitivity achievable in brighter QSOs. Second, the Mg {\sc ii} data at 
low $z$ and the C {\sc iv} data at high $z$ come from many of the same QSO
spectra.  If observers systematically overestimated or underestimated the 
signal-to-noise ratios, the effect should show up in both Mg {\sc ii} and 
C {\sc iv}, which it does not. Finally, the largest homogeneous sample within 
our full sample shows the same effect observed in the full sample, despite the 
smaller range in M$_{\rm V}$ and the smaller total sample size.

Could the effect be intrinsic to the QSOs? The radiation fields near QSOs
are thought to affect the ionization state and reduce the counts of the
nearby absorbers (Ellingson et al.\ 1994).
However, this effect is opposite to the correlation between C {\sc iv}
absorber counts and M$_{\rm V}$ seen here.  In the results above, we have
excluded the regions within 5000 km/s of the QSOs, and the results do not
change significantly when regions up to 20000 km/s from the QSOs are
excluded. Unless ejection velocities of absorbers from QSOs often exceed
several tens of thousands of km/s {\it and} narrow line profiles are
maintained, our results are unlikely to be due to an effect intrinsic to QSOs.
Furthermore, an ejection related explanation would have to involve an
excess of absorbers related to luminosity and a luminosity independent
intergalactic sample of absorbers. To explain the identical
$d{\cal N}/d{\sl W}$ slopes in Fig.\ 2 for C {\sc iv}, the ejected
sample and the intergalactic sample would have to have virtually identical
distributions of $d{\cal N}/d{\sl W}$.  This possibility seems unlikely,
given the different physical origins.  Finally, some ionization effect would
be expected, for which we find no evidence in our discussion of Si {\sc ii}
data below.

A final possibility remains, namely that mass indirectly associated with the C
{\sc iv} absorbers brightens the QSOs through gravitational lensing. However,
the lensing is probably not of the kind that produces multiple QSO images for
two reasons. First, the HST QSO Snapshot survey (Maoz et al.\, 1993) and
numerous ground-based optical and radio surveys for multiply-imaged QSOs
(e.g.\ Jaunsen et al.\, 1995) have shown that the fraction of bright
QSOs with resolved multiple images within several arcseconds is much less
than 1\%. For multi-image lensing to explain our results, the image
splittings would have to be either less than the resolution of these surveys
or so much greater that their identifications have been overlooked. Clusters
are able to produce multiple QSO images with several arcminute separations,
but only for fairly specific mass configurations (Paczy\'{n}ski \& Gorski
1981) which are probably not common enough (Turner, Ostriker, \& Gott 1984) to
explain the amplitude of our results.  Second, the probability density for
encountering lenses producing multiple images is quite small near the QSOs
(Turner,Ostriker, \& Gott 1984). Figure 5 shows the region of space
sampled by the catalog used here by showing the distances of the absorbers
and the distances of the background QSOs.  Various selection criteria
discussed earlier are delimited by dashed lines.  Any lenses in our 
sample must be at a redshift $z \gtrsim 1.5$ and 
within several hundred comoving Mpc of the QSOs to be
associated with detected C {\sc iv} absorbers. Impact parameters through
clusters of  galaxies do not have to be restricted to those that produce
multiple images in order to have significant amplification, and as the
QSO-lens distance becomes smaller, the probability of producing only a single
image increases (as can be seen in Figure 6 of Young et al.\ 1980). If the
probability density for this type of lensing is greatest within a few hundred
Mpc of a QSO, we would not expect a significant correlation between Mg {\sc
ii} absorber counts and QSO luminosity, since most of the Mg {\sc ii}
absorbers are found much farther away from their QSOs (Figure 5).
Unfortunately the sample of Mg {\sc ii} absorbers within a few hundred Mpc of
their QSOs is still too small to see if their numbers are correlated with QSO
luminosity.

Differences between the lensing signatures of Mg {\sc ii} and C {\sc iv} could
also arise if there is a segregation of absorber ions in how they trace
lensing potentials, or in how the ionization environments differ near
the QSO compared with the average over all space.  We have analyzed the
Si {\sc ii} 1527{\AA} (rest wavelength) absorbers which have ionization
properties similar to Mg {\sc ii}, but are found in the space regime (Fig.\
5) of C {\sc iv} (rest wavelengths 1548{\AA}, 1550{\AA}). The counts versus
M$_{\rm V}$ distribution of Si {\sc ii}, shown in Figure 4, is more similar to
that of C {\sc iv}, than of Mg {\sc ii}.  The ${\chi}^{2}$
probability that ${\cal N}_{obs}/{\cal N}_{exp}$ vs.\ M$_{\rm V}$
distribution for Si {\sc ii} is consistent with that for Mg {\sc ii} is
0.43 while the probability is 0.94 for C {\sc iv}. This result is consistent
with the idea that the lack of a count/luminosity correlation in the Mg
{\sc ii} sample is due to a reduced probability of intercepting a lens far
away from a QSO,  rather than to, say, an ionization effect in QSO ejecta
that leaves C {\sc iv} but eliminates Mg {\sc ii}, or to some effect of
segregation of C {\sc iv}, Mg {\sc ii}, and Si {\sc ii} near the QSO.

It will be important to obtain samples in a given absorber species over the
largest redshift range possible, either by extending the redshift range
for C {\sc iv}, Si {\sc ii}, Si {\sc iv}, Al {\sc ii}, and O {\sc i} to lower
redshifts by deblending the listed lines from the Ly $\alpha$ forest
(Kulkarni et al. 1996), or deblending features from telluric absorption and
atmospheric emission at long wavelengths to obtain higher redshift samples of
Mg {\sc ii} (Caulet 1989).  Large homogeneous QSO and absorber surveys, such
as the Sloan Digital Sky Survey, are being planned, which should increase the
total QSO and absorber samples at all redshifts by a factor exceeding 100.
Another survey is under way to
obtain a large number of absorbers in a several complete samples of
selectively faint QSOs (Borra et al.\ 1996).  We expect that the
mean C {\sc iv} $d{\cal N}/dz$ in this sample will be significantly lower than
the average $d{\cal N}/dz$ in Fig.\ 1. 

Further work will also be needed to see if detailed lensing models are
consistent with our results.

\section{Conclusions}
The published sample of intervening C {\sc iv} absorbers in QSO spectra
shows a strong
luminosity dependent number density excess compared to the published
Mg {\sc ii} absorber sample.  Systematic selection effects in the data have
been tested for by searching for differences in the absorber equivalent width
distributions in intrinsically bright and faint QSO samples, and by checking
that the effect appears in the largest homogeneous sample within our
heterogeneous but carefully treated sample from the literature.  No
explanation of this sort was found.  While we do not believe that the effect
shown here is due to a selection bias, such as overestimating the equivalent
width detection limit for faint QSOs relative to bright ones, if it were,
then all existing sample of QSO absorbers
are subject to this effect, and all work on dN/dz for CIV must be reconsidered
in light of this bias.  The reduced catalog we used is available
on line from DEVB, so these types of effects can be examined by others.

Furthermore, no evidence was found that more luminous QSOs have gas with
very high ejection velocities, though it cannot be ruled out entirely.  Since
we undertook this study to search for a lensing bias in the counts of QSO
absorbers, we briefly explored the types of lensing that might be relevant. 
Various observational selection effects allow the C {\sc iv} absorbers to
sample only the space at $z \gtrsim 1.5$ and 
within 400 Mpc of the background QSOs, whereas the Mg
{\sc ii} data set does not sample this region.  Therefore, relevant
lensing must occur at $z \gtrsim 1.5$ and 
within 400 Mpc of the QSOs.  It must consist of
amplification without multiple images, as the latter are not seen in the the
vast majority of QSOs.

Several observational tests were mentioned that may shed light on the cause
of the C {\sc iv} excess in more luminous QSOs.  These involve achievable
increases in sample size and in redshift coverage for the Mg {\sc ii} and 
C {\sc iv} samples.

The effect, no matter what the cause, is so strong that $d{\cal N}/dz$ --
an important statistic for inferring morphology of the absorbers --
must be considered uncertain until the effect is understood.

\acknowledgments
The authors would like to acknowledge helpful discussions with Jim Gunn,
Daniel Holz, Arieh K\"{o}nigl, Don Lamb, Bohdan Paczy\'{n}ski, and Bob Wald. 
DGY acknowledges an early discussion with Craig Hogan that led to this search
for lensing. DEVB was supported by the Adler Fellowship at the University of
Chicago, and by the NASA Space Telescope grant GO-06007.01-94A.  JMQ is
supported by the Compton Fellowship -- NASA grant GDP93-08.

\bigskip
\noindent
Electronic mail: \\
\medskip
devb@oddjob.uchicago.edu (DEVB) \\
\medskip
jmq@oddjob.uchicago.edu (JMQ) \\
\medskip
don@oddjob.uchicago.edu (DGY) \\
\medskip
yanny@sdss.fnal.gov (BY) \\

\newpage
\figcaption[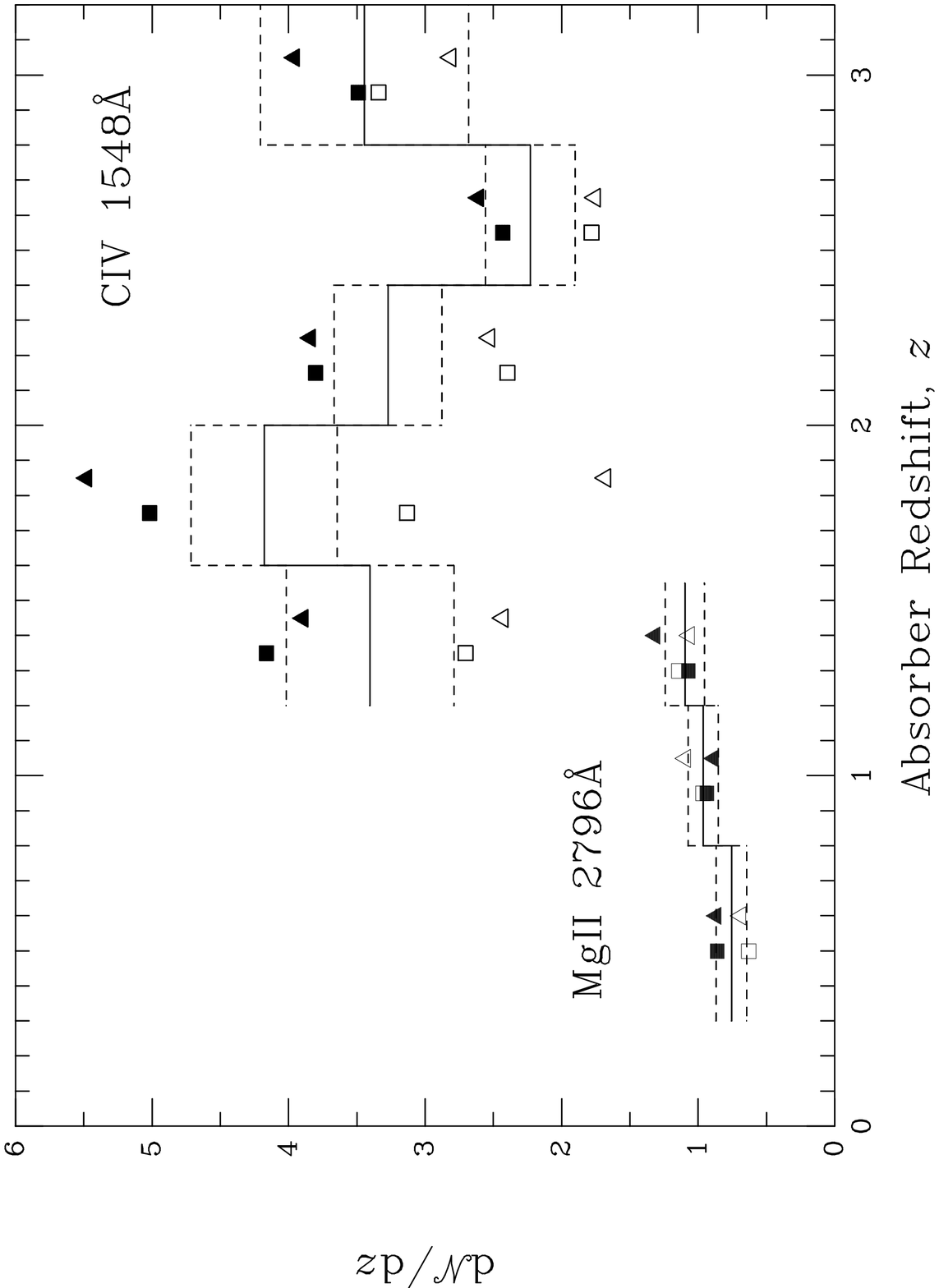]{The redshift number density of absorbers for Mg {\sc ii}
($0.3<z<1.55$) and C {\sc iv} ($1.2<z<3.2$), in bright (filled symbols) and
faint (open symbols) QSOs.  Squares show the QSO division by the median
absolute magnitude, M$_{\rm V}$, and triangles show the division by a 
line fit to apparent
magnitude versus redshift.  The histograms show the average $d{\cal N}/dz$
and 1$\sigma$ deviation for 100 Monte Carlo simulations.}

\figcaption[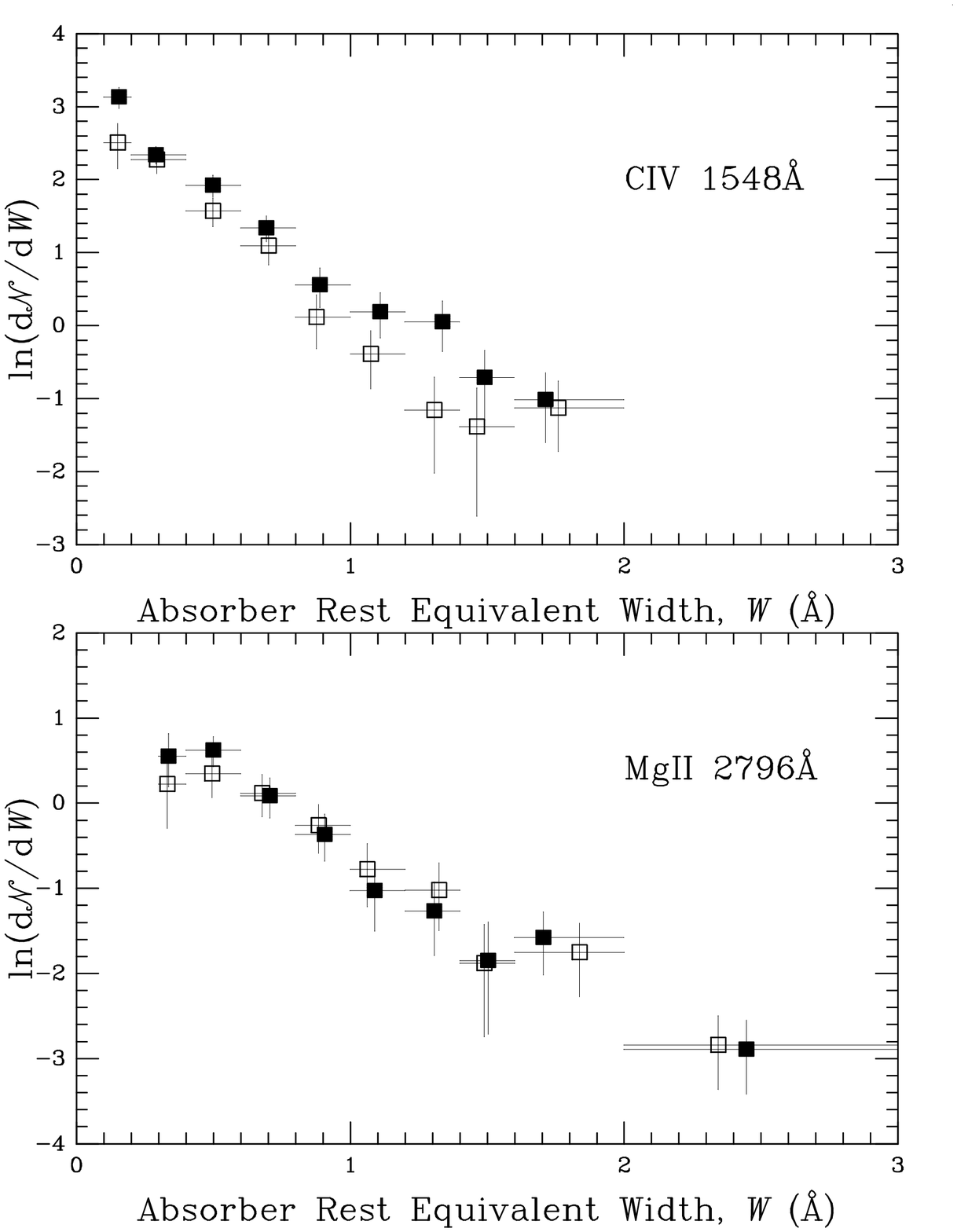]{The rest equivalent width number density of absorbers
for C {\sc iv} and Mg {\sc ii}.  Filled squares show the results for bright
QSOs, and open squares for faint QSOs, separated by their median M$_{\rm V}$.}

\figcaption[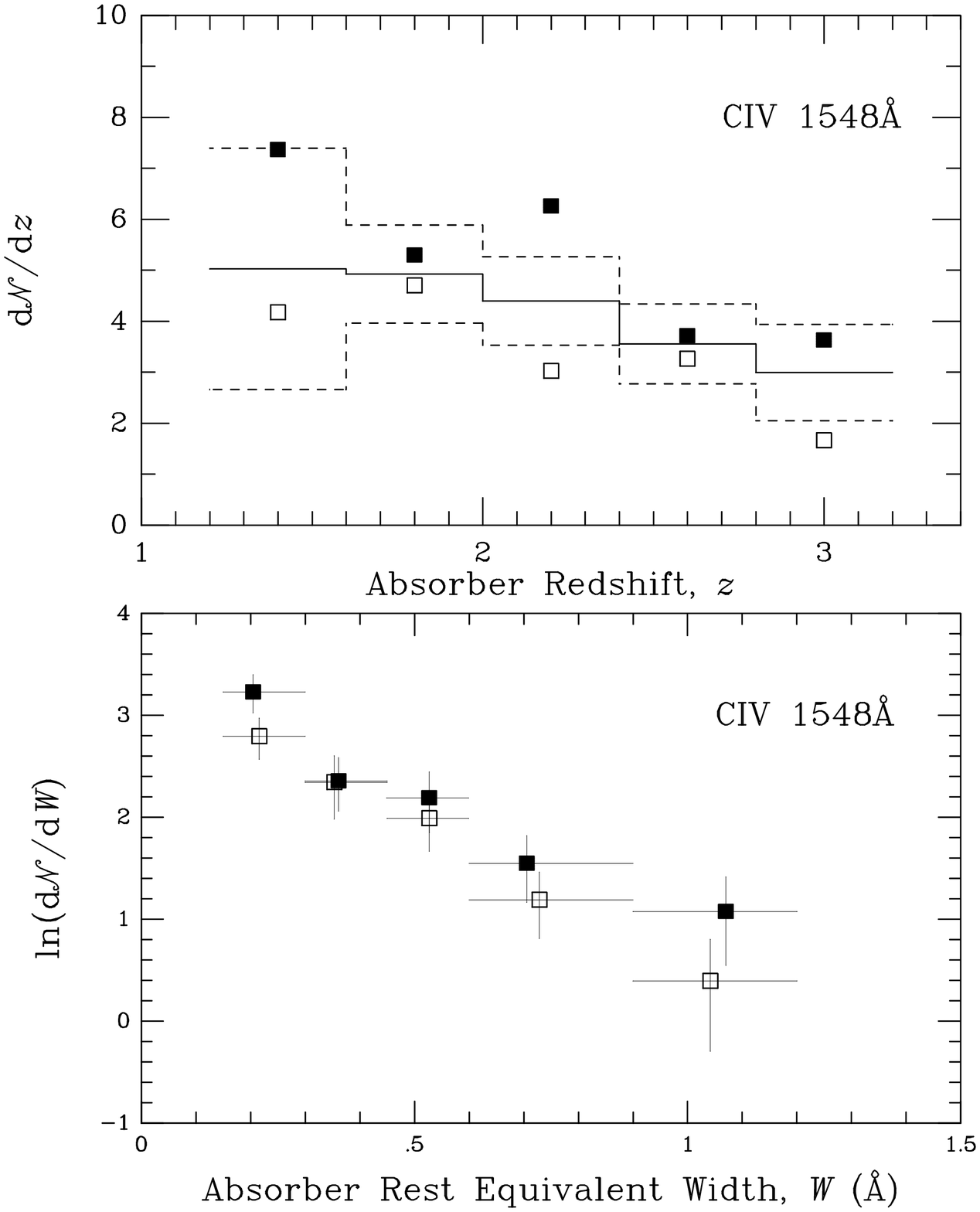]{The C {\sc iv} redshift number density (top) and 
equivalent width number density (bottom) for bright (filled squares) and
faint (open squares) QSOs in the homogeneous sample of
Sargent, Boksenberg, \& Steidel (1988), and Steidel (1990).  
The histograms in the
top plot are as in Fig.\ 1. Only absorbers with
$1.2 < z < 3.2$ and ${\sl W} > 0.15${\AA} ($5 \sigma$) have been used.  
Absorbers lying within 150 km/s have been combined (as in Steidel 1990) for the
$d{\cal N}/dz$ plot.}

\figcaption[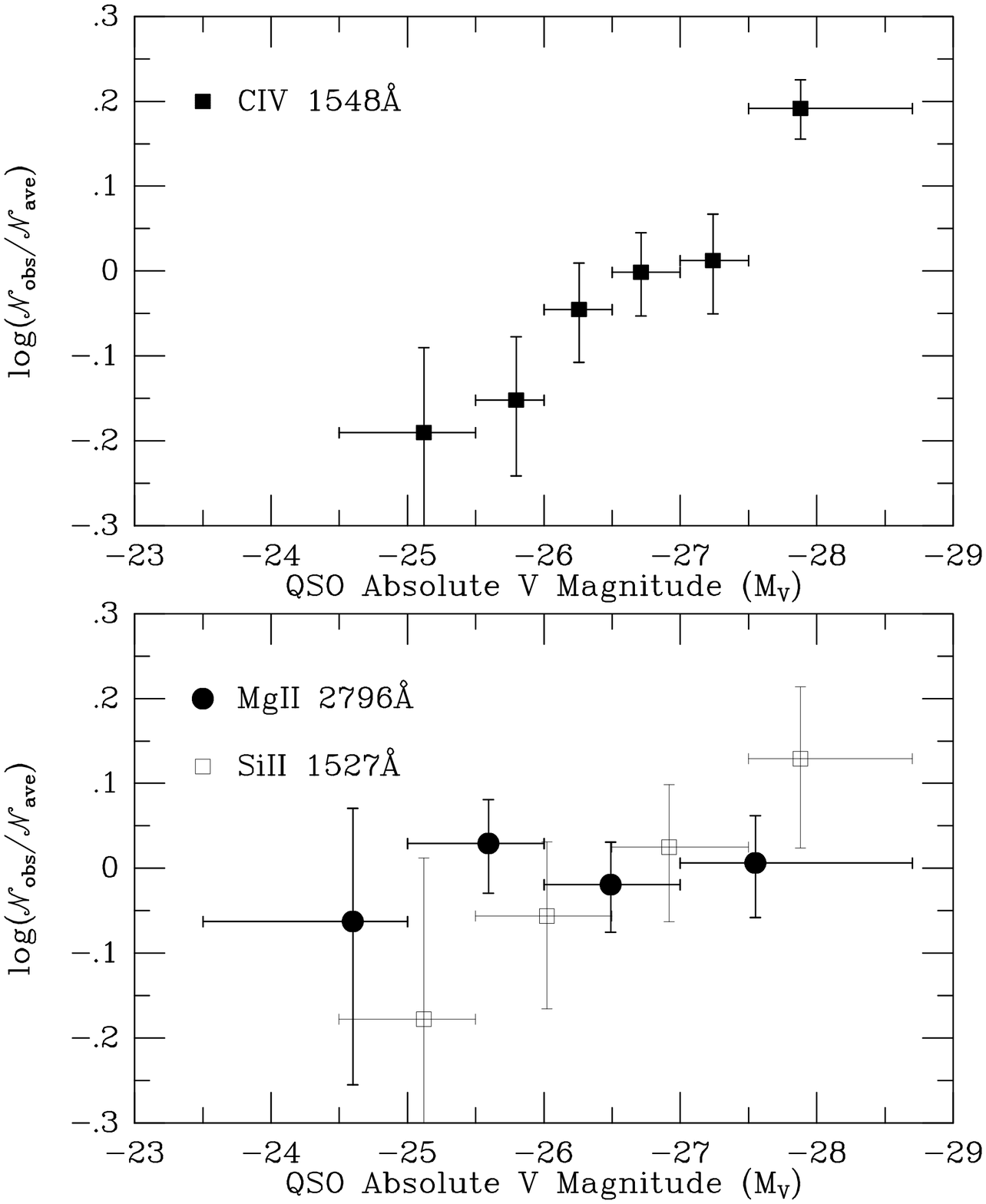]{The ratio of the number of observed absorbers ${\cal
N}_{\rm obs}$, to the number of absorbers expected on average ${\cal N}_{\rm
ave}$, as a function of QSO absolute magnitude.  There is a strong increasing
trend in the C {\sc iv} sample (filled squares), while there is no
significant trend in the Mg {\sc ii} sample (filled circles). The Si {\sc ii}
sample (open squares) also shows an increasing trend.}

\figcaption[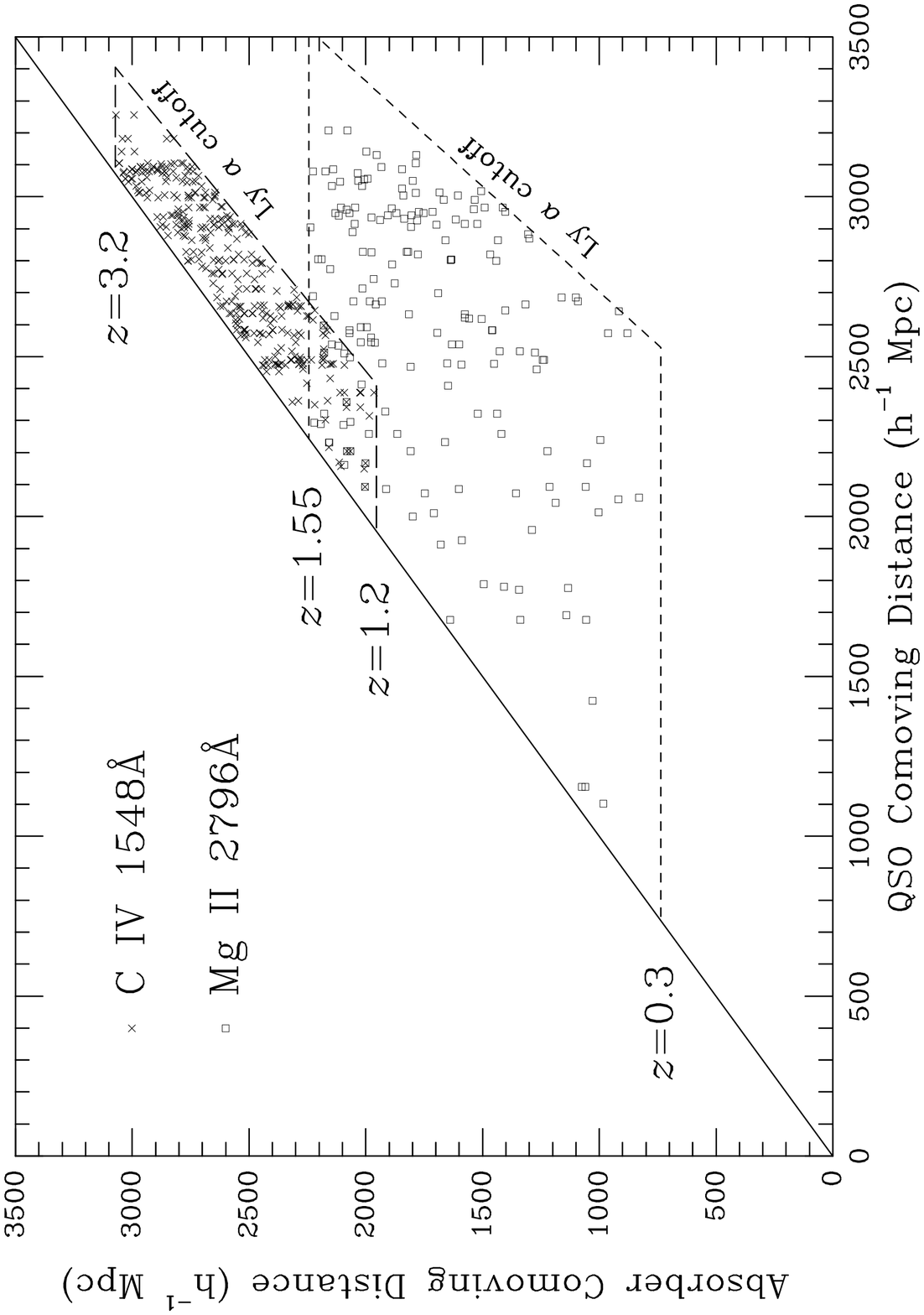]{The comoving distances to the QSOs and absorbers in the
C {\sc iv} and Mg {\sc ii} samples.  The regions inside the dashed lines are
bounded by the atmospheric redshift limits and Ly $\alpha$ forest cutoff for
each absorber ion, and show where the absorbers could have been detected.}

\newpage
\begin{figure}
\figurenum{1}
\plotone{f1.eps}
\caption{}
\end{figure}
\newpage
\begin{figure}
\figurenum{2}
\plotone{f2.eps}
\caption{}
\end{figure}
\newpage
\begin{figure}
\figurenum{3}
\plotone{f3.eps}
\caption{}
\end{figure}
\newpage
\begin{figure}
\figurenum{4}
\plotone{f4.eps}
\caption{}
\end{figure}
\newpage
\begin{figure}
\figurenum{5}
\plotone{f5.eps}
\caption{}
\end{figure}

\end{document}